\documentclass[twocolumn,aps,prl,showpacs,preprintnumbers,amsmath,amssymb,floatfix]{revtex4-1}
\usepackage[T1]{fontenc}
\usepackage[utf8]{inputenc}
\usepackage{xcolor}
\usepackage{graphicx}
\usepackage[unicode=true,
 bookmarks=false,
 breaklinks=false,pdfborder={0 0 1},backref=section,colorlinks=true]
 {hyperref}
\hypersetup{
 linkcolor=blue,urlcolor=blue,citecolor=blue,anchorcolor=blue}

\makeatletter

\newcommand*\LyXThinSpace{\,\hspace{0pt}}

\makeatother

\begin{document}
\bibliographystyle{apsrev4-1}
\title{Quantum Coherence as a Thermodynamic Resource Beyond the Classical
Uncertainty Bound}
\author{Shanhe Su\textsuperscript{1}}
\email{sushanhe@xmu.edu.cn;}

\author{Cong Fu\textsuperscript{1}}
\author{Ousi Pan\textsuperscript{1}}
\author{Shihao Xia\textsuperscript{1}}
\author{Fei Liu\textsuperscript{2}}
\email{feiliu@buaa.edu.cn}

\author{Jincan Chen\textsuperscript{1}}
\email{jcchen@xmu.edu.cn;}

\affiliation{\textsuperscript{1}Department of Physics, Xiamen University, Xiamen,
361005, People's Republic of China}
\affiliation{\textsuperscript{2}School of Physics, Beihang University, Beijing
100083, China}
\date{\today}
\begin{abstract}
Thermodynamic uncertainty relations (TURs) establish a fundamental
trade-off between current precision and entropy production in nonequilibrium
systems, yet the role of genuine quantum coherence in these bounds
remains unresolved. Here we develop a general framework that explicitly
incorporates quantum coherence into TURs for Markovian open quantum
systems. By combining the quantum Cramér--Rao inequality with a Dyson
series expansion of the parametrically deformed Lindblad dynamics,
we derive a generalized TUR containing a coherence-sensitive correction
that naturally separates classical and quantum contributions. The
coherent term originates from the off-diagonal components of the nonequilibrium
steady-state density matrix and systematically relaxes the classical
precision--dissipation bound, allowing enhanced current precision
without additional entropy production. Applying our theory to a three-level
quantum maser, we demonstrate that steady-state coherence enables
precision beyond the classical limit over a broad parameter regime
while remaining consistent with the generalized bound. Our results
identify quantum coherence as a genuine thermodynamic resource and
provide a unified framework bridging classical and quantum uncertainty
relations.
\end{abstract}
\maketitle
\textit{Introduction.}---The thermodynamic uncertainty relation (TUR)
stands as a key principle of modern thermodynamics, unifying dissipation,
fluctuations, and precision in small-scale systems far from equilibrium
\cite{Barato2015,Horowitz2020,Miller2021,Koyuk2020,Macieszczak2018,Hasegawa2021}.
It formalizes a fundamental trade-off in which achieving high precision
in physical currents, such as particle transport, heat conduction,
or work extraction, requires a correspondingly large rate of entropy
production, reflecting the irreversibility inherent in nonequilibrium
processes. In its classical steady-state formulation \cite{Barato2015,Pietzonka17,Koyuk2022,Chun2019,Koyuk2021},
the TUR is expressed as
\begin{equation}
\frac{D}{J^{2}}\geq\frac{2}{\sigma},
\end{equation}
where $J$ denotes the mean current, $D$ characterizes the fluctuations
of the current, and $\sigma$ represents the entropy production rate.
Throughout this work, we adopt natural units with $\hbar=k_{B}=1$.
This inequality implies that the relative uncertainty of any current
cannot be made arbitrarily small without incurring a thermodynamic
cost. In practice, the TUR imposes stringent limits on the energetic
efficiency of biomolecular machines, nanoscale electronic devices,
and artificial molecular engines, making it a key principle for understanding
the operation of systems at the mesoscale \cite{Koyuk2021,Brandner2025,Tajima2021,Dieball2023,Yoshimura2021,Hasegawa2019,Hasegawa2021-1,Pietzonka2022,Hartich2021}.

Despite their broad applicability, existing TURs are largely restricted
to classical or incoherent quantum regimes, leaving a fundamental
question unresolved: what role do genuinely quantum effects play in
the cost--precision trade-off? In particular, coherence---a uniquely
quantum resource absent in classical systems---can strongly influence
energy and charge transport \cite{Dong2012,Li2015,Rodrigues2019,Kwon2018,Su2018,Su2016}.
Its precise impact, however, remains unclear: does coherence merely
introduce additional noise, tightening the uncertainty bound, or can
it be leveraged to surpass classical precision limits without incurring
extra thermodynamic cost? Addressing this question is crucial not
only for developing a comprehensive framework of quantum thermodynamics
but also for guiding the design of quantum devices that harness coherence
to achieve enhanced performance.

Explicit quantum models, such as the thermoelectric generators \cite{Ptaszy=000144ski2018}
and three-level masers \cite{Kalaee2021,Singh2023}, reveal that quantum
coherence can overcome the constraints imposed by the power, fluctuations,
and entropy production. This insight has motivated a broad search
for general quantum extensions of thermodynamic bounds \cite{Rignon-Bret2021,Cangemi2020,Menczel2021}.
A quantum thermodynamic uncertainty relation for arbitrary continuous
measurements has been derived from the quantum Cramér--Rao inequality,
in which the quantum Fisher information is naturally separated into
contributions from dynamical activity and coherent dynamics \cite{Hasegawa2020}.
Finite-time uncertainty bounds, applicable to arbitrary initial states
and operation durations, reveal that quantum coherence fundamentally
limits precision by amplifying the trade-off between the fluctuations
and the energetic cost in generic dissipative systems \cite{Vu2022}.
A fundamental link has recently been established between the kinetic
uncertainty relation and the quantum coherence, where the coherence
factor quantifies the incompatibility between the steady-state density
matrix and the system Hamiltonian, and the quantum Fisher information
scales directly with the dynamical activity \cite{Prech2025}.

Although quantum coherence can be characterized by various measures
\cite{Vu2022-1,Francica2020,Shiraishi2025,Camati2016}, the absence
of a transparent and physically intuitive quantifier that connects
coherence to TURs remains a major unresolved problem. In contrast,
within the framework of kinetic uncertainty relation, the role of
coherence has been explicitly identified and quantitatively characterized,
revealing how quantum superpositions can tighten or relax classical
kinetic uncertainty relation \cite{Prech2025}. For TURs, however,
such a clear correspondence is still missing. Establishing a coherence-based
measure that directly captures the quantum contribution to TURs would
not only deepen our understanding of how coherence alters the trade-off
among the precision, dissipation, and entropy production, but also
clarify whether coherence can be regarded as a fundamental thermodynamic
resource. Such a development could ultimately lead to a unified picture
of uncertainty relations across classical and quantum regimes, where
coherence plays a central role in determining the achievable limits
of nonequilibrium performance.

In this work, we develop a general theoretical framework that explicitly
incorporates quantum coherence into TURs. Drawing on the recent progresses
in quantum thermodynamics and information geometry, we introduce a
coherence-based quantifier that captures the intrinsic quantum contribution
to the trade-off between precision, dissipation, and entropy production.
We demonstrate that quantum coherence can qualitatively modify---and
in certain regimes, relax---the classical TUR bound. These findings
establish coherence as a genuine thermodynamic resource that governs
the ultimate precision limits in nonequilibrium quantum systems, thereby
providing a unified perspective that reconciles quantum formulations
of uncertainty relations.

\textit{Thermodynamic uncertainty relation with quantum coherence\label{sec:setup}}---To
make these ideas concrete, we now formulate the TUR in the presence
of quantum coherence. By explicitly incorporating coherence contributions,
we demonstrate that coherence can fundamentally reshape the trade-off
between the current precision and the thermodynamic cost. In particular,
we derive the generalized bound

\begin{equation}
\frac{D}{J^{2}}\geq\frac{2(1-\psi_{\mathbf{\mathrm{cl}}}-\psi_{\mathrm{qc}})^{2}}{\sigma}.
\end{equation}
where the dimensionless quantities $\psi_{\mathbf{\mathrm{cl}}}$
and $\psi_{\mathbf{\mathrm{qc}}}$ quantify, respectively, the classical
and quantum-coherent contributions to the lower bound. 

Physically,$\psi_{\mathbf{\mathrm{cl}}}$ arises from classical fluctuations
in the nonequilibrium dynamics, while $\psi_{\mathbf{\mathrm{qc}}}$
captures the additional effect of coherence originating from superpositions
in the energy eigenbasis or coherent driving. The presence of $\psi_{\mathbf{\mathrm{qc}}}$
can either tighten or loosen the bound, implying that quantum coherence
may enhance precision. This generalized TUR thus provides a unified
framework for assessing current fluctuations in coherent quantum thermodynamics.

\textit{General model\label{sec:model}}---We consider a Markovian
open quantum system that is simultaneously coupled to multiple thermal
reservoirs maintained at different temperatures. The dynamics of the
system are described by the time-dependent density matrix $\rho$,
which evolves according to the Lindblad master equation \cite{Breuer2001,Rivas2012,Alicki2010,Liu2018,Lindblad1976,Alicki1979}:

\begin{equation}
\frac{d}{dt}\rho=\mathcal{L}\rho=-i\left[H,\rho\right]+\sum_{k}\mathcal{D}\left(L_{k},\rho\right).
\end{equation}
Here, $\mathcal{L}$ denotes the Lindblad generator, $H$ is the time-independent
Hamiltonian of the system, $\left[\cdot,\cdot\right]$ represents
the commutator, $L_{k}$ are the Lindblad jump operators associated
with different dissipative channels, and the dissipator is defined
as $\mathcal{D}\left(L_{k},\rho\right)=L_{k}\rho L^{\dagger}_{k}-\frac{1}{2}\left\{ L^{\dagger}_{k}L_{k},\rho\right\} $
with $\left\{ \cdot,\cdot\right\} $ being the anticommutator. Since
the system is simultaneously coupled to multiple reservoirs at different
temperatures, the resulting dissipative dynamics generally drive it
toward a nonequilibrium steady state $\rho_{s}$, characterized by
the stationarity condition $\mathcal{L}\rho_{s}=0$. 

As shown in Ref. \cite{Prech2025}, for the quantum jump unraveling,
the mean current $J$ and the noise $D$ admit explicit expressions:
\begin{equation}
J=\operatorname{Tr}\{\mathcal{J}\rho\}
\end{equation}
and
\begin{equation}
D=\sum_{k}\operatorname{Tr}\left\{ \nu^{2}_{k}L_{k}\rho L^{\dagger}_{k}\right\} -2\operatorname{Tr}\left\{ \mathcal{J}\mathcal{L}^{+}\mathcal{J}\rho\right\} ,
\end{equation}
respectively. Here, $\mathcal{J}\rho=\sum_{k}\nu_{k}L_{k}\rho L^{\dagger}_{k}$
with $\nu_{k}$ being the weight associated with the contribution
of the jump operator $L_{k}$, and $\mathcal{L}^{+}$ denotes the
Drazin pseudoinverse of the Lindblad generator $\mathcal{L}$. The
superoperator $\mathcal{J}$ serves as the current superoperator. 

We now derive the TUR using the quantum Cramér--Rao inequality. In
Refs. \cite{Hasegawa2020,Vu2022,Prech2025,Liu2018}, the Cramér--Rao
inequality was employed to derive quantum TURs by introducing virtual
perturbations to the system dynamics through suitable modifications
of the Hamiltonian and jump operators. Motivated by these works, and
in order to establish the TUR in Eq. (2), we introduce a controlled
deformation of the jump operators in the Lindblad master equation
{[}Eq. (3){]}:
\begin{equation}
L_{k}\rightarrow L_{k,\theta}=\sqrt{1+\ell_{k}(t)\theta}L_{k},
\end{equation}
where $\theta$ serves as a small, dimensionless deformation parameter.
The coefficient 
\begin{equation}
\ell_{k}(t)=\frac{\operatorname{tr}\left\{ L_{k}\rho_{\mathrm{s}}L^{\dagger}_{k}\right\} -\operatorname{tr}\left\{ L_{k^{\prime}}\rho_{\mathrm{s}}L^{\dagger}_{k^{\prime}}\right\} }{\operatorname{tr}\left\{ L_{k}\rho_{\mathrm{s}}L^{\dagger}_{k}\right\} +\operatorname{tr}\left\{ L_{k^{\prime}}\rho_{\mathrm{s}}L^{\dagger}_{k^{\prime}}\right\} }
\end{equation}
acts as a time-dependent weighting function associated with the $k$-th
dissipative channel, and satisfies the relation $\ell_{k}(t)=-\ell_{k^{\prime}}(t)$
\cite{Vu2022,Vu2022-1,Vu2021,Vu2021-1}. Here, $L_{k^{\prime}}$ denotes
the jump operator corresponding to the time-reversed process of the
$k$th jump. We assume the local detailed balance $L_{k}=e^{\Delta s_{k}/2}L^{\dagger}_{k^{\prime}}$,
where $\Delta s_{k}$ denotes the entropy change of the environment
associated with the jump. This condition is satisfied in most physical
scenarios \cite{Breuer2001,Kosloff2017,Funo2019,Esposito2009}.

Under this deformation, the Lindblad master equation becomes
\begin{equation}
\frac{d}{dt}\rho=\mathcal{L}_{\theta}\rho=-i\left[H,\rho\right]+\sum_{k}\left[1+\ell_{k}(t)\theta\right]\mathcal{D}\left(L_{k},\rho\right).
\end{equation}
By design, the original dynamics is recovered in the limit $\theta\rightarrow0$,
i.e., $\mathcal{L}_{\theta}\rightarrow\mathcal{L}$. From a physical
perspective, this deformation can be understood as introducing a virtual
bias between the paired jump processes. This perturbation serves as
a controlled tool to explore the sensitivity of observable fluctuations
to small changes in the jump rates. 

By invoking the generalized quantum Cramér--Rao bound \cite{Hasegawa2020,Vu2022,Helstrom1976,Ito2020,Supp},
we have
\begin{equation}
\frac{Var_{\theta}[N(\tau)]_{\theta=0}}{\left(\partial_{\theta}\mathrm{\mathbb{E}}_{\theta}[N(\tau)]_{\theta=0}\right)^{2}}\geq\frac{1}{\mathcal{I}(0)},
\end{equation}
where $N(\tau)$ denotes the integrated current (or counting observable)
over the time interval $\left[0,\tau\right]$. $Var_{\theta}[N(\tau)]$
represents the variance of $N(\tau)$, while $\mathrm{\mathbb{E}}_{\theta}[N(\tau)]$
is the expectation value of $N(\tau)$ under the dynamics parameterized
by $\theta$. In the steady state, we have $Var_{\theta}[N(\tau)]_{\theta=0}=D\tau$
and $\mathrm{\mathbb{E}}_{\theta}[N(\tau)]_{\theta=0}=J\tau$ (Supplemental
Material). The derivative term $\partial_{\theta}\mathrm{E}_{\theta}[N(\tau)]$
characterizes the sensitivity of the mean current to perturbations
in $\theta$. It can be explicitly evaluated as (Supplemental Material)

\begin{equation}
\partial_{\theta}\mathrm{\mathbb{E}}_{\theta}[N(\tau)]_{\theta=0}=\tau J(1-\psi_{\mathbf{\mathrm{cl}}}-\psi_{\mathrm{qc}}).
\end{equation}
In the steady state, the factor $\psi_{\mathbf{\mathrm{cl}}}=A_{\mathrm{cl}}/J$
quantifies the classical contribution present, while $\psi_{\mathbf{\mathrm{qc}}}=A_{\mathrm{qc}}/J$
captures the contribution arising from quantum coherence. Here, $A_{\mathrm{cl}}=\sum_{k}\nu_{k}\langle1|L^{\ast}_{k}\otimes L_{k}\mathcal{L}^{+}\mathcal{D_{\mathit{l}}}|\rho_{\mathrm{diag}}\rangle$
and $A_{\mathrm{qc}}=\sum_{k}\nu_{k}\langle1|L^{\ast}_{k}\otimes L_{k}\mathcal{L}^{+}\mathcal{D_{\mathit{l}}}|\rho_{\mathrm{off}}\rangle$.
In the vectorization formalism, the density operator $\rho_{\mathrm{s}}$
is mapped to the vector $|\rho_{\mathrm{s}}\rangle=\sum_{i,j}\rho_{s,ij}|i\rangle\otimes|j\rangle$,
where $\left\{ |i\rangle\right\} $ is the orthonormal basis of the
system Hamiltonian. The diagonal part $|\rho_{\mathrm{diag}}\rangle=\sum_{i}\rho_{s,ii}|i\rangle\otimes|i\rangle$
corresponds to populations, while the off-diagonal part $|\rho_{\mathrm{off}}\rangle=\sum_{i,j}\rho_{s,ij}|i\rangle\otimes|j\rangle$
( $i\neq j$ ) represents quantum coherence. The quantity $A_{\mathrm{qc}}$
is associated with the coherence part $|\rho_{\mathrm{off}}\rangle$
\cite{Tajima2021}. In the absence of quantum coherence, $|\rho_{\mathrm{off}}\rangle=0$
and consequently $A_{\mathrm{qc}}=0$, consistent with the physical
expectation that any quantum advantage due to coherence vanishes when
no quantum superposition is present.

We note that Eq. (10) is consistent with that reported in Ref. \cite{Vu2025}.
In Ref. \cite{Vu2025}, the result was derived via a perturbative
expansion of the parametrized Caldeira--Leggett--Saito--Lindblad
(CKSL) equation. Here, we employ the Dyson series solution to obtain
the same expression (Supplemental Material). As discussed in Ref.
\cite{Vu2025}, in the classical limit, the term $\psi_{\mathbf{\mathrm{cl}}}+\psi_{\mathrm{qc}}$
in Eq. (10) vanishes, which means that it contains information related
to quantum coherence. Therefore, to investigate the influence of quantum
coherence on TUR, we introduce a diagonalized state in the energy
eigenbasis and separate out the factor $\psi_{\mathrm{qc}}$ associated
with quantum coherence from this term.

The quantity $\mathcal{I}(0)$ denotes the quantum Fisher information
evaluated at $\theta=0$, computed following the result of Gammelmark
and Mølmer \cite{Prech2025,Gammelmark2014}. We find that $\mathcal{I}(0)$
is upper bounded by the entropy production rate (see also Supplemental
Material for a detailed derivation), namely,
\begin{equation}
\mathcal{I}(0)\leq\frac{1}{2}\tau\sigma,
\end{equation}
where the steady-state entropy production rate is given by $\sigma=\sum_{k}\operatorname{tr}\left\{ L^{\dagger}_{k}L_{k}\rho_{\mathrm{s}}\right\} \Delta s_{k}$
\cite{Tajima2021,Vu2022}. Combining Eqs. (8) and (10) with the quantum
Cramér--Rao bound {[}Eq. (7){]} directly yields the TUR with quantum
coherence {[}Eq. (2){]}.

\textit{Three level maser \label{sec:maser}}---As an illustrative
case, we analyze the main results using a quantum Scovil--Schulz-DuBois
(SSDB) three-level maser operating as a heat engine {[}see Fig. 1(a){]}
\cite{Boukobza2006,Boukobza2007,Kalaee2021-1,Boukobza2006-1,Boukobza2013}.
The system interacts with a classical electric field and is coupled
to both a hot and a cold heat bath, characterized by inverse temperatures
$\beta_{\alpha}(\alpha=h,c)$. The three energy eigenstates of the
system are denoted by $\left|h\right\rangle $, $\left|c\right\rangle $
and $\left|x\right\rangle $. Bath $\alpha$, with population $n_{\alpha}$,
induces transitions between state $\left|\alpha\right\rangle $ and
$\left|x\right\rangle $ at rate $\gamma_{\alpha}$. An external ac
field couples states $\left|h\right\rangle $ and $\left|c\right\rangle $
with strength $\Omega$ and frequency $\omega_{d}=\Delta+\omega_{c}-\omega_{h}$,
where $\Delta$ is a detuning parameter.

In one operational cycle, the system is excited by the hot bath $h$
into the excited state $x$, from which it relaxes to $\left|c\right\rangle $
while emitting a photon into the cold bath $c$. The cycle closes
with the emission of a photon into the driving field, thereby producing
work. To ensure the validity of the local master equation, we restrict
our analysis to the weak-driving regime\cite{Geva1996}. In a rotating
frame, the evolution of the density matrix is governed by Eq. (3)
(see Supplemental Material \cite{Supp} for details) with the Hamiltonian
$H=-\Delta\sigma_{cc}+\Omega\left(\sigma_{ch}+\sigma_{hc}\right)$
and the jump operators $L_{1}=\sqrt{\gamma_{h}n_{h}}\sigma_{xh},L_{2}=\sqrt{\gamma_{h}\left(n_{h}+1\right)}\sigma_{hx},L_{3}=\sqrt{\gamma_{c}n_{c}}\sigma_{xc}$,
and $L_{4}=\sqrt{\gamma_{c}\left(n_{c}+1\right)}\sigma_{cx}$, where
the transition operators are defined as $\sigma_{ij}=\left|i\right\rangle \left\langle j\right|$. 

\begin{figure}
\includegraphics[scale=0.3]{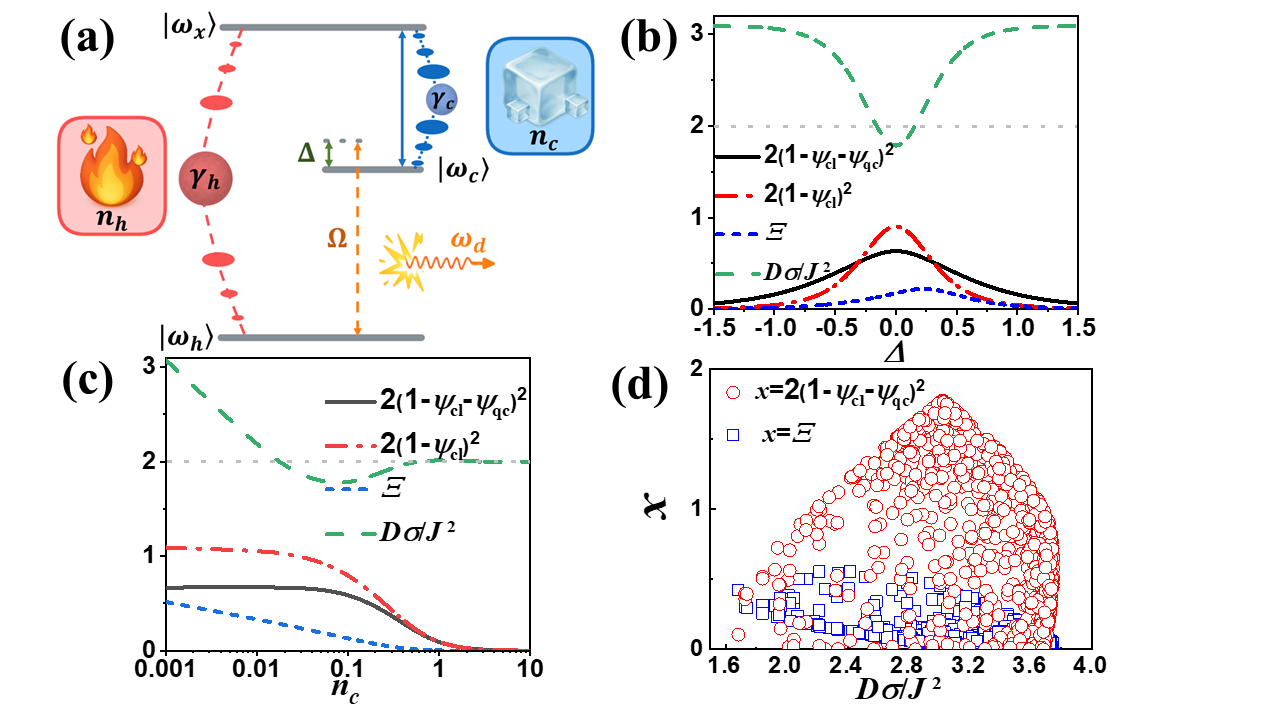}\caption{(a) Schematic diagram of the quantum Scovil--Schulz-DuBois (SSDB)
three-level maser. The thermodynamic uncertainty relation (TUR) $D\sigma/J^{2}$
(dashed green line) for the quantum SSDB three-level maser, the bound
$2(1-\psi_{\mathbf{\mathrm{cl}}})^{2}$ (dash-dotted red line) which
neglects the coherence factor $\psi_{\mathrm{qc}}$, the coherence
bound $2(1-\psi_{\mathbf{\mathrm{cl}}}-\psi_{\mathrm{qc}})^{2}$ (solid
black line), and the quantum TUR bound $\mathcal{\mathrm{\Xi}\mathrm{=\frac{\sigma}{\Upsilon+\Psi}}}$
(short-dashed blue line) are plotted as functions of (a) the detuning
$\Delta$ with $n_{c}=0.06$ and $\Omega=0.25$, and (c) the bath
population $n_{c}$ with $\Delta=0$ and $\Omega=0.25$. (d) The scatter
plots of $2(1-\psi_{\mathbf{\mathrm{cl}}}-\psi_{\mathrm{qc}})^{2}$(red
circles) and $\mathrm{\Xi}$ (blue cubes) against $D\sigma/J^{2}$,
where $\Delta$ is randomly chosen from -1.5 to 1.5, $\Omega$ is
randomly chosen from 0.01 to 0.8, and $n_{c}=0.027$. The remaining
parameters are fixed at $\gamma_{c}=2,\gamma_{h}=0.1$, and $n_{h}=5$. }
\end{figure}

Adopting Hasegawa’s framework, the quantum TUR for the SSDB in the
steady state takes the form $D\sigma/J^{2}\geq\sigma/\left(\Upsilon+\Psi\right)=\mathcal{\mathrm{\Xi}}$
\cite{Hasegawa2020,Prech2025}. Here, $\Upsilon=\sum_{k}L_{k}\rho_{s}L^{\dagger}_{k}$
denotes the quantum dynamical activity, while the expression of the
coherent contribution $\Psi=-4\operatorname{Tr}\left[\mathcal{K}_{1}\mathcal{L}^{\mathrm{+}}\mathcal{K}_{2}\left(\rho_{s}\right)+\mathcal{K}_{2}\mathcal{L}^{\mathrm{+}}\mathcal{K}_{1}\left(\rho_{s}\right)\right]$.
The superoperators $\mathcal{K}_{1}$ and $\mathcal{K}_{2}$ are defined
as $\mathcal{K}_{1}\left(\rho\right)=-\mathrm{i}H\rho+\frac{1}{2}\sum_{k}\left(L_{k}\rho L^{\dagger}_{k}-L^{\dagger}_{k}L_{k}\rho\right)$,
and $\mathcal{K}_{2}\left(\rho\right)=\mathrm{i}\rho H+\frac{1}{2}\sum_{k}\left(L_{k}\rho L^{\dagger}_{k}-\rho L^{\dagger}_{k}L_{k}\right)$. 

Figure 1(b) shows the curves of the exact TUR ratio $D\sigma/J^{2}$,
the classical bound $2\left(1-\psi_{\mathrm{cl}}\right)^{2}$ which
neglects coherence, our new coherence-sensitive bound $2\left(1-\psi_{\mathrm{cl}}-\psi_{\mathrm{qc}}\right)^{2}$,
and Hasegawa's quantum bound $\Xi=\sigma/(\Upsilon+\Psi)$ varying
with the detuning $\Delta$. In the vicinity of resonance ($\Delta\rightarrow0$),
where coherent driving generates maximal steady-state coherence between
the two lower levels, the classical TUR is strongly violated and the
exact precision falls below 2, i.e., $D\sigma/J^{2}\leq2$. Crucially,
our coherence-corrected bound remains below the exact ratio, i.e.,
$D\sigma/J^{2}\geq2(1-\psi_{{\rm cl}}-\psi_{{\rm qc}})^{2}$.

This conclusion is further reinforced in Fig. 1(c), where the cold-bath
population $n_{c}$ is varied. As $n_{c}$ increases, the nonequilibrium
drive weakens, suppressing both the particle current and the steady-state
coherence. Consequently, the coherence correction $\psi_{{\rm qc}}$
diminishes and the lower bound $2(1-\psi_{{\rm cl}}-\psi_{{\rm qc}})^{2}$
continuously approaches the incoherent limit $2(1-\psi_{{\rm cl}})^{2}$.
Throughout this entire range, the exact TUR ratio $D\sigma/J^{2}$
remains above our bound, confirming the universal validity of Eq.
(2). The dependence on both $\Delta$ and $n_{c}$ indicates that
$\psi_{{\rm qc}}$ quantitatively captures the coherence-induced relaxation
of the thermodynamic cost for precision.

To further assess the robustness of the theory, we randomly sample
a broad parameter space by varying both the detuning $\Delta$ and
the driving strength $\Omega$ . As illustrated in Fig. 1(d), our
coherence-sensitive bound (red circles) lies strictly below the exact
TUR ratio, confirming its universal validity. More importantly, it
consistently provides a significantly tighter lower bound than the
existing quantum TUR $\Xi$ (blue cubes) across the majority of the
parameter regime. 

Taken together, Figs. 1(b)--1(d) establish two central results. First,
quantum coherence systematically relaxes the classical precision-dissipation
trade-off while preserving a universal uncertainty relation. Second,
the coherence-sensitive correction yields a significantly tighter
lower bound than existing quantum TURs, demonstrating that the coherent
component of the response is the missing ingredient required to quantitatively
characterize nonequilibrium precision beyond the classical limit.

\textit{Conclusions.---}We have developed a general theory of thermodynamic
uncertainty relations in the presence of quantum coherence. Combining
the quantum Cramér--Rao inequality with a Dyson-series expansion
of deformed Lindblad dynamics, we identified a coherence-sensitive
correction that reveals how steady-state coherence reshapes the precision--dissipation
trade-off. Our framework establishes quantum coherence as a genuine
thermodynamic resource and bridges classical and quantum uncertainty
relations, providing a foundation for understanding precision limits
in nonequilibrium quantum systems.\\

\textit{Acknowledgement.---}We thank Professor Tan Van Vu for valuable
discussions. This work has been supported by the Natural Science Foundation
of Fujian Province (2023J01006), National Natural Science Foundation
of China (12364008 and 12365006), and Fundamental Research Fund for
the Central Universities of China (20720240145).

\end{document}


\title{Supplemental material for ``Quantum Coherence as a Thermodynamic
Resource Beyond the Classical Uncertainty Bound''}
\author{Shanhe Su\textsuperscript{1}}
\email{sushanhe@xmu.edu.cn;}

\author{Cong Fu\textsuperscript{1}}
\author{Ousi Pan\textsuperscript{1}}
\author{Shihao Xia\textsuperscript{1}}
\author{Jincan Chen\textsuperscript{1}}
\author{Fei Liu\textsuperscript{2}}
\email{feiliu@buaa.edu.cn}

\affiliation{\textsuperscript{1}Department of Physics, Xiamen University, Xiamen,
361005, People's Republic of China}
\affiliation{\textsuperscript{3}School of Physics, Beihang University, Beijing
100083, China}
\begin{abstract}
The supplementary information is organized as follows. The upper bound
of the quantum Fisher information is derived in Section I, with its
dependence on the entropy production rate being elucidated. In Section
II, an analytical evaluation of the derivative $\partial_{\theta}E_{\theta}[N(\tau)]$
employing the Dyson series expansion is presented, which is crucial
for quantifying quantum coherence. In Section III, we provide the
derivation of the thermodynamic uncertainty relations (TURs) in the
presence of quantum coherence, highlighting how coherence modifies
the classical bounds. Explicit expressions for the mean current $J$
and the noise $D$ are given in Section IV, which are used to characterize
the precision\textendash dissipation trade-off. The master equation
for the quantum Scovil\textendash Schulz-DuBois (SSDB) three-level
maser is analyzed in Section V.
\end{abstract}
\maketitle

\section{the upper bound of the quantum Fisher information}

The quantum Fisher information at the steady state is evaluated by
using the formalism developed by Gammelmark and Mølmer \cite{Gammelmark2014}
\begin{equation}
\mathcal{I}(\theta)=4\tau\sum_{k}\operatorname{Tr}\left\{ \left(\partial_{\theta}L_{k,\theta}\right)^{\dagger}\left(\partial_{\theta}L_{k,\theta}\right)\rho_{\mathrm{s}}\right\} -4\tau\left(\operatorname{Tr}\left\{ \mathcal{L}_{L}\mathcal{L}^{+}\mathcal{L}_{R}\rho_{\mathrm{s}}\right\} +\operatorname{Tr}\left\{ \mathcal{L}_{R}\mathcal{L}^{+}\mathcal{L}_{L}\rho_{\mathrm{s}}\right\} \right),
\end{equation}
where
\begin{equation}
\mathcal{L}_{L}\rho=-i\partial_{\theta}H\rho-\frac{1}{2}\sum_{k}\partial_{\theta}\left(L^{\dagger}_{k,\theta}L_{k,\theta}\right)\rho+\sum_{k}\left(\partial_{\theta}L_{k,\theta}\right)\rho L^{\dagger}_{k,\theta}\text{,}
\end{equation}
and
\begin{equation}
\mathcal{L}_{R}\rho=i\rho\partial_{\theta}H-\frac{1}{2}\sum_{k}\rho\partial_{\theta}\left(L^{\dagger}_{k,\theta}L_{k,\theta}\right)+\sum_{k}L_{k,\theta}\rho\left(\partial_{\theta}L_{k,\theta}\right)^{\dagger}.
\end{equation}

An auxiliary dynamics parameterized by the parameter $\theta$ is
considered as follows:

\begin{equation}
H_{\theta}=H,L_{k,\theta}(t)=\sqrt{1+\ell_{k}(t)\theta}L_{k},
\end{equation}
where
\begin{equation}
\ell_{k}=\frac{\operatorname{tr}\left\{ L_{k}\rho_{\mathrm{s}}L^{\dagger}_{k}\right\} -\operatorname{tr}\left\{ L_{k^{\prime}}\rho_{\mathrm{s}}L^{\dagger}_{k^{\prime}}\right\} }{\operatorname{tr}\left\{ L_{k}\rho_{\mathrm{s}}L^{\dagger}_{k}\right\} +\operatorname{tr}\left\{ L_{k^{\prime}}\rho_{\mathrm{s}}L^{\dagger}_{k^{\prime}}\right\} }.
\end{equation}

In the analysis of thermal dissipation dynamics, we assume the local
detailed balance condition $L_{k}=e^{\Delta s_{k}/2}L^{\dagger}_{k^{\prime}}$,
which holds in most physically relevant cases \cite{Vu2022}. Here,
the operator $L^{\dagger}_{k^{\prime}}$ represents the reversed jump
of the $k$th jump, and $\Delta s_{k}$ denotes the corresponding
entropy change in the environment induced by this jump.

For $\theta\rightarrow0$, Eqs. (2) and (3) reduce to
\begin{equation}
\mathcal{L}_{L}\rho=\frac{1}{2}\sum_{k}\left(\ell^{2}_{k}L_{k}\rho L^{\dagger}_{k}-\ell^{2}_{k}L^{\dagger}_{k}L_{k}\rho\right),
\end{equation}
\begin{equation}
\mathcal{L}_{R}\rho=\frac{1}{2}\sum_{k}\left(\ell^{2}_{k}L_{k}\rho L^{\dagger}_{k}-\ell^{2}_{k}\rho L^{\dagger}_{k}L_{k}\right).
\end{equation}
These operators are traceless, i.e., $Tr\left\{ \mathcal{L}_{L}\rho\right\} =0$
and $Tr\left\{ \mathcal{L}_{R}\rho\right\} =0$, becasue of the cyclic
property of the trace. Furthermore, the action of $\mathcal{L}^{+}$
preserves the trace property of the superoperator. Consequently, the
last two terms in Eq. (1) vanish, and in the limit $\theta\rightarrow0$,
the quantum Fisher information (QFI) reduces to
\begin{align}
\mathcal{I}(0) & =\tau\sum_{k}\ell^{2}_{k}\operatorname{Tr}\left\{ L^{\dagger}_{k}L_{k}\rho_{\mathrm{s}}\right\} \nonumber \\
 & =\tau\sum_{k}\ell^{2}_{k}\operatorname{Tr}\left\{ L_{k}\rho_{\mathrm{s}}L^{\dagger}_{k}\right\} \\
 & =\frac{1}{2}\tau\sum_{k}\ell^{2}_{k}\left(\operatorname{tr}\left\{ L_{k}\rho_{\mathrm{s}}L^{\dagger}_{k}\right\} +\operatorname{tr}\left\{ L_{k^{\prime}}\rho_{\mathrm{s}}L^{\dagger}_{k^{\prime}}\right\} \right)\nonumber \\
 & =\frac{1}{2}\tau\sum_{k}\frac{\left(\operatorname{tr}\left\{ L_{k}\rho_{\mathrm{s}}L^{\dagger}_{k}\right\} -\operatorname{tr}\left\{ L_{k^{\prime}}\rho_{\mathrm{s}}L^{\dagger}_{k^{\prime}}\right\} \right)^{2}}{\operatorname{tr}\left\{ L_{k}\rho_{\mathrm{s}}L^{\dagger}_{k}\right\} +\operatorname{tr}\left\{ L_{k^{\prime}}\rho_{\mathrm{s}}L^{\dagger}_{k^{\prime}}\right\} }\nonumber \\
 & \leq\frac{1}{2}\tau\sigma.
\end{align}
In the final inequality, we have used the rigorous lower bound on
the entropy production rate
\begin{equation}
\sigma\geq\sum_{k}\frac{\left(\operatorname{tr}\left\{ L_{k}\rho_{\mathrm{s}}L^{\dagger}_{k}\right\} -\operatorname{tr}\left\{ L_{k^{\prime}}\rho_{\mathrm{s}}L^{\dagger}_{k^{\prime}}\right\} \right)^{2}}{\operatorname{tr}\left\{ L_{k}\rho_{\mathrm{s}}L^{\dagger}_{k}\right\} +\operatorname{tr}\left\{ L_{k^{\prime}}\rho_{\mathrm{s}}L^{\dagger}_{k^{\prime}}\right\} }
\end{equation}
derived in Refs. \cite{Vu2022,Tajima2021}. 

\section{Analytical Evaluation of the Derivative $\partial_{\theta}\mathbb{E}_{\theta}[N(\tau)]$}

When $\theta=0$, the expected value of the time integrated current
$N(\tau)$ is given by
\begin{equation}
\begin{aligned}\mathrm{\mathbb{E}}[N(\tau)] & =\sum_{k}\int^{\tau}_{0}dt\nu_{k}\operatorname{Tr}\left\{ L_{k}\rho_{t}L^{\dagger}_{k}\right\} \\
 & =\sum_{k}\int^{\tau}_{0}dt\nu_{k}\operatorname{Tr}\left\{ L^{\dagger}_{k}L_{k}e^{\mathcal{L}t}\rho_{0}\right\} ,
\end{aligned}
\end{equation}
where $\rho_{0}$ denotes the initial state. Given the modified jump
operators in Eq. (4), the expectation value of $N(\tau)$ as a function
of the parameter $\theta$ is expressed as
\begin{equation}
\mathrm{\mathbb{E}}_{\theta}[N(\tau)]=\sum_{k}\int^{\tau}_{0}dt\nu_{k}\left(1+\ell_{k}\theta\right)\operatorname{Tr}\left\{ L^{\dagger}_{k}L_{k}e^{\mathcal{L}_{\theta}t}\rho_{0}\right\} .
\end{equation}

Taking the partial derivative of $\mathrm{E}_{\theta}[N(\tau)]$ with
respect to $\theta$, we obtain
\begin{equation}
\begin{aligned}\partial_{\theta}\mathrm{\mathbb{E}}_{\theta}[N(\tau)]= & \sum_{k}\int^{\tau}_{0}dt\nu_{k}\ell_{k}\operatorname{Tr}\left\{ L^{\dagger}_{k}L_{k}e^{\mathcal{L}_{\theta}t}\rho_{0}\right\} \\
 & +\sum_{k}\int^{\tau}_{0}dt\nu_{k}\left(1+\ell_{k}\theta\right)\operatorname{Tr}\left\{ L^{\dagger}_{k}L_{k}\left(\partial_{\theta}e^{\mathcal{L}_{\theta}t}\right)\rho_{0}\right\} .
\end{aligned}
\end{equation}

We emphasize that the initial state $\rho_{0}$ remains independent
of the parameter $\theta$. Recall that our analysis focuses on the
limit $\theta\rightarrow0$ and noting that $\ell_{k}=-\ell_{k^{\prime}}$,
$\nu_{k}=-\nu_{k^{\prime}}$, and $\nu_{k}\ell_{k}=\nu_{k^{\prime}}\ell_{k^{\prime}}$.
Accordingly, if $\rho_{0}$ is taken to be the steady state $\rho_{\mathrm{s}}$,
the first term in Eq. (11) is simplified as follows:
\begin{align}
\lim_{\theta\rightarrow0}\sum_{k}\int^{\tau}_{0}dt\nu_{k}\ell_{k}\operatorname{Tr}\left\{ L^{\dagger}_{k}L_{k}e^{\mathcal{L}_{\theta}t}\rho_{0}\right\}  & =\sum_{k}\int^{\tau}_{0}dt\nu_{k}\ell_{k}\operatorname{Tr}\left\{ L^{\dagger}_{k}L_{k}\rho_{\mathrm{s}}\right\} \nonumber \\
 & =\frac{1}{2}\sum_{k}\int^{\tau}_{0}dt\nu_{k}\ell_{k}\left(\operatorname{Tr}\left\{ L^{\dagger}_{k}L_{k}\rho_{\mathrm{s}}\right\} +\operatorname{Tr}\left\{ L^{\dagger}_{k^{\prime}}L_{k^{\prime}}\rho_{\mathrm{s}}\right\} \right)\nonumber \\
 & =\frac{1}{2}\sum_{k}\int^{\tau}_{0}dt\nu_{k}\ell_{k}\left(\operatorname{Tr}\left\{ L_{k}\rho_{\mathrm{s}}L^{\dagger}_{k}\right\} +\operatorname{Tr}\left\{ L_{k^{\prime}}\rho_{\mathrm{s}}L^{\dagger}_{k^{\prime}}\right\} \right)\nonumber \\
 & =\frac{1}{2}\sum_{k}\int^{\tau}_{0}dt\nu_{k}\left(\operatorname{tr}\left\{ L_{k}\rho_{\mathrm{s}}L^{\dagger}_{k}\right\} -\operatorname{tr}\left\{ L_{k^{\prime}}\rho_{\mathrm{s}}L^{\dagger}_{k^{\prime}}\right\} \right)\nonumber \\
 & =\sum_{k}\int^{\tau}_{0}dt\nu_{k}\operatorname{tr}\left\{ L_{k}\rho_{\mathrm{s}}L^{\dagger}_{k}\right\} \nonumber \\
 & =\tau J,
\end{align}
where $J=\operatorname{tr}\left\{ \mathcal{J}\rho_{\mathrm{s}}\right\} $
is the average of the stochastic current with and the current superoperator
$\mathcal{J}$ acts as $\mathcal{J}\rho=\sum_{k}\nu_{k}L_{k}\rho L^{\dagger}_{k}$.
In the fourth equality, the definition of $\ell_{k}$ given in Eq.
(5) has been used.

To simplify the second term in Eq. (13), we employ the Dyson series
solution for the evolution operator in the interaction picture. Recall
that $\mathcal{L}_{\theta}=\mathcal{L}+\theta\mathcal{D_{\mathit{l}}}$,
where $\mathcal{D_{\mathit{l}}\mathrm{\rho=\mathit{\sum_{k}l_{k}\left(L_{k}\rho L^{\dagger}_{k}-\mathrm{\frac{1}{2}}\left\{ L^{\dagger}_{k}L_{k},\rho\right\} \right)}}}$
\cite{Prech2025}. We consider the first-order expansion of $e^{\mathcal{L}_{\theta}t}$
with respect to the small parameter $\theta$. By introducing the
free-evolution operator $U_{0}(t)=e^{\mathcal{L}t}$, we define the
evolution operator in the interaction picture $U_{I}(t;\theta)$ through
the relation
\begin{equation}
e^{\mathcal{L}_{\theta}t}=U_{0}(t)U_{I}(t;\theta),
\end{equation}
which implies $U_{I}(t;\theta)=e^{-\mathcal{L}t}e^{\mathcal{L}_{\theta}t}$.
Differentiating both sides with respect to $t$ gives 
\begin{equation}
\mathcal{L}_{\theta}U_{0}U_{I}=\mathcal{L}U_{0}U_{I}+U_{0}\frac{d}{dt}U_{I}.
\end{equation}

Comparing both sides, the common term $\mathcal{L}U_{0}U_{I}$ cancels,
leaving

\begin{equation}
U_{0}\frac{dU_{I}}{dt}=\theta\mathcal{D_{\mathit{l}}}U_{0}U_{I}.
\end{equation}
Then, multiplying both sides by $U_{0}(t)^{-1}=e^{-\mathcal{L}t}$,
we obtain
\begin{equation}
\frac{dU_{I}}{dt}=\theta e^{-\mathcal{L}t}\mathcal{\mathcal{D_{\mathit{l}}}}e^{\mathcal{L}t}U_{I}=\theta\mathcal{D}_{I}(t)U_{I}.
\end{equation}
where $\mathcal{D}_{I}(t)=e^{-\mathcal{L}t}D_{\mathit{l}}e^{\mathcal{L}t}$
is the perturbation operator in the interaction picture. The formal
solution of this equation is given by the time-ordered exponential
\begin{equation}
U_{I}=\mathcal{T}\exp\left(\theta\int^{\tau}_{0}\mathcal{D}_{I}(s)ds\right),
\end{equation}
where $\mathcal{T}$ denotes the time-ordering operator, and $U_{I}(0;\theta)=I$.
Expanding this expression yields the Dyson series
\begin{equation}
U_{I}=I+\theta\int^{\tau}_{0}\mathcal{D}_{I}\left(t_{1}\right)dt_{1}+\theta^{2}\int^{\tau}_{0}dt_{1}\int^{t_{1}}_{0}dt_{2}\mathcal{D}_{I}\left(t_{1}\right)\mathcal{D}_{I}\left(t_{2}\right)+\cdots.
\end{equation}

To first order in $\theta$, this reduces to
\begin{equation}
U_{I}=I+\theta\int^{\tau}_{0}\mathcal{D}_{I}\left(t_{1}\right)dt_{1}+\mathcal{O}\left(\theta^{2}\right)=I+\theta\int^{\tau}_{0}e^{-\mathcal{L}t_{1}}D_{\mathit{l}}e^{\mathcal{L}t_{1}}dt_{1}+\mathcal{O}\left(\theta^{2}\right).
\end{equation}

Returning to the original picture, we have
\begin{align}
e^{\mathcal{L}_{\theta}t} & =U_{0}U_{I}\nonumber \\
 & =e^{\mathcal{L}t}\left[I+\theta\int^{\tau}_{0}e^{-\mathcal{L}t_{1}}D_{\mathit{l}}e^{\mathcal{L}t_{1}}dt_{1}+\mathcal{O}\left(\theta^{2}\right)\right]\nonumber \\
 & =e^{\mathcal{L}t}+\theta\int^{t}_{0}dt_{1}e^{\mathcal{L}\left(t-t_{1}\right)}\mathcal{D_{\mathit{l}}}e^{\mathcal{L}t_{1}}+\mathcal{O}\left(\theta^{2}\right)
\end{align}

Substituting this expression into Eq. (13), and using Eq. (14) together
with the fact that $e^{\mathcal{L}t_{1}}\rho_{\mathrm{s}}=\rho_{\mathrm{s}}$
(the steady state), we find
\begin{equation}
\partial_{\theta}\mathrm{\mathbb{E}}_{\theta}[N(\tau)]_{\theta=0}=\tau J+\sum_{k}\nu_{k}\int^{\tau}_{0}dt\int^{t}_{0}dt_{1}\operatorname{Tr}\left\{ L^{\dagger}_{k}L_{k}e^{\mathcal{L}\left(t-t_{1}\right)}\mathcal{D_{\mathit{l}}}\rho_{\mathrm{s}}\right\} .
\end{equation}

To facilitate the evaluation of the second term on the right-hand
side, it is convenient to adopt a vectorized representation \cite{Prech2025}.
Within this framework, the trace and the steady state correspond to
the left and right eigenvectors of the Liouvillian associated with
the zero eigenvalue, and are, respectively, given by
\begin{equation}
\langle1|\mathcal{L}=0\quad\text{ and }\quad\mathcal{L}|\rho\rangle=0.
\end{equation}

As shown by Prech \cite{Prech2025}, in the long-time limit, the double
integral in Eq. (15) can be approximated as
\begin{equation}
\int^{\tau}_{0}dt\int^{t}_{0}dt_{1}e^{\mathcal{L}\left(t-t_{1}\right)}=\frac{\tau^{2}}{2}|\rho\rangle\langle1|-\tau\mathcal{L}^{+},
\end{equation}
where $\mathcal{L}^{+}$ denotes the Drazin inverse of the Liouvillian
$\mathcal{L}$. Since $\mathcal{D_{\mathit{l}}}$ is traceless, the
$\tau^{2}$-proportional term vanishes upon substitution of Eq. (25)
into Eq. (23), yielding:
\begin{equation}
\partial_{\theta}\mathrm{\mathbb{E}}_{\theta}[N(\tau)]_{\theta=0}=\tau J-\tau\sum_{k}\nu_{k}\langle1|L^{\ast}_{k}\otimes L_{k}\mathcal{L}^{+}\mathcal{D_{\mathit{l}}}|\rho_{\mathrm{s}}\rangle.
\end{equation}
We note that this result is consistent with that reported in Ref.
\cite{Vu2025}. In Ref. \cite{Vu2025}, the result was derived via
a perturbative expansion of the parametrized Caldeira\textendash Leggett\textendash Saito\textendash Lindblad
(CKSL) equation. Here, we employ the Dyson series and the long-time
limit approximation in Eq. (25) to obtain the same expression. As
discussed in Ref. \cite{Vu2025}, in the classical limit, the second
term in Eq. (26) vanishes, which means that it contains information
related to quantum coherence. Therefore, to investigate the influence
of quantum coherence on TUR, we will subsequently introduce a diagonalized
state in the energy eigenbasis and separate out the factor associated
with quantum coherence from this term.

In the vectorization formalism, the density operator $\rho_{\mathrm{s}}$
is mapped to a vector $|\rho_{\mathrm{s}}\rangle=\sum_{i,j}\rho_{s,ij}|i\rangle\otimes|j\rangle$,
where $\left\{ |i\rangle\right\} $ is the orthonormal basis of the
system Hamiltonian. Here, the diagonal part $|\rho_{\mathrm{diag}}\rangle=\sum_{i}\rho_{s,ii}|i\rangle\otimes|i\rangle$
correspond to populations, while the off-diagonal part $|\rho_{\mathrm{off}}\rangle=\sum_{i,j}\rho_{s,ij}|i\rangle\otimes|j\rangle$
( $i\neq j$ ) represent quantum coherences. From this, we have the
equality
\begin{align}
\sum_{k}\nu_{k}\langle1|L^{\ast}_{k}\otimes L_{k}\mathcal{L}^{+}\mathcal{D_{\mathit{l}}}|\rho_{\mathrm{s}}\rangle & =\sum_{k}\nu_{k}\langle1|L^{\ast}_{k}\otimes L_{k}\mathcal{L}^{+}\mathcal{D_{\mathit{l}}}|\rho_{\mathrm{diag}}\rangle+\sum_{k}\nu_{k}\langle1|L^{\ast}_{k}\otimes L_{k}\mathcal{L}^{+}\mathcal{D_{\mathit{l}}}|\rho_{\mathrm{off}}\rangle\nonumber \\
 & =A_{\mathrm{cl}}+A_{\mathrm{qc}}
\end{align}
where $A_{\mathrm{cl}}=\sum_{k}\nu_{k}\langle1|L^{\ast}_{k}\otimes L_{k}\mathcal{L}^{+}\mathcal{D_{\mathit{l}}}|\rho_{\mathrm{diag}}\rangle$
and $A_{\mathrm{qc}}=\sum_{k}\nu_{k}\langle1|L^{\ast}_{k}\otimes L_{k}\mathcal{L}^{+}\mathcal{D_{\mathit{l}}}|\rho_{\mathrm{off}}\rangle$.
The quantity $A_{\mathrm{qc}}$ is associated with coherence part
$|\rho_{\mathrm{off}}\rangle$ \cite{Tajima2021}. If the system\textquoteright s
state contains no quantum coherence, then the vector $|\rho_{\mathrm{off}}\rangle=0$,
and $A_{\mathrm{qc}}$ is zero. This aligns with the physical expectation
that any quantum advantage due to coherence would vanish in the absence
of quantum superposition.

From Eqs. (26) and (27), it follows that
\begin{equation}
\partial_{\theta}\mathrm{\mathbb{E}}_{\theta}[N(\tau)]_{\theta=0}=\tau J(1-\psi_{\mathbf{\mathrm{cl}}}-\psi_{\mathrm{qc}}),
\end{equation}
where the factor $\psi_{\mathbf{\mathrm{cl}}}=A_{\mathrm{cl}}/J$
quantifies the classical contribution present in the steady state,
and $\psi_{\mathbf{\mathrm{qc}}}=A_{\mathrm{qc}}/J$ quantifies the
contribution from quantum coherence.

\section{Derivation of Thermodynamic Uncertainty Relations with Quantum Coherence}

The following inequality is guaranteed by the quantum Cramér\textendash Rao
bound \cite{Hasegawa2020,Helstrom1976} :
\begin{equation}
\frac{Var_{\theta}[N(\tau)]_{\theta=0}}{\left(\partial_{\theta}\mathrm{\mathbb{E}}_{\theta}[N(\tau)]_{\theta=0}\right)^{2}}\geq\frac{1}{\mathcal{I}(0)}.
\end{equation}
The left-hand side (LHS) quantifies the ratio of fluctuations (noise)
to the squared sensitivity of the mean current, reflecting a trade-off
between signal and noise, while the right-hand side (RHS) gives the
fundamental lower limit set by the quantum Fisher information (QFI).
Combining this inequality with Eqs. (9) and (30) leads to
\begin{equation}
\frac{D}{J^{2}}\geq\frac{2(1-\psi_{\mathbf{\mathrm{cl}}}-\psi_{\mathrm{qc}})^{2}}{\sigma},
\end{equation}
where the scaled variance (referred to as the noise) of the stochastic
current is defined as $D=\frac{\operatorname{Var}[N(\tau)]}{\tau}$.
Eq. (30) can be viewed as a generalization of the conventional TUR,
which is bounded by the entropy production rate $\sigma$ and the
coherence factor $\psi_{\mathrm{qc}}$. 

\section{The calculations of the mean current $J$ and the noise $D$ }

Full counting statistics (FCS) provides a framework to characterize
the complete probability distribution of transported quantities\textemdash such
as charge, energy, particle number, work, or heat\textemdash in quantum
or mesoscopic systems. In what follows, we adopt the approach outlined
in Refs. \cite{Prech2025,Schaller2013}, to introduce the concept
of counting fields and the associated cumulant generating function.
The cumulant generating function not only encodes the full statistical
information about the transport process but also provides a systematic
way to extract cumulants of arbitrary order. Subsequently, we employ
the method presented in Ref. \cite{Landi2024}, which develops a powerful
formalism for deriving exact analytical expressions of the mean and
variance in the long-time limit. This technique is particularly valuable
because it enables a clear separation between the deterministic trends
and the fluctuation-induced effects, thereby revealing universal features
of transport statistics across a wide range of quantum systems.

Let $\rho(N,t)$ denote the density matrix of the system conditioned
on $N$ photons being transferred to bath $c$ during a time $t$.
The corresponding master equation can be written as
\begin{equation}
\dot{\rho}(N,t)=\mathcal{L}_{0}\rho(N,t)+\mathcal{L}_{-}\rho(N+1,t)+\mathcal{L}_{+}\rho(N-1,t),
\end{equation}
where the operator $\mathcal{L}_{+(-)}$ describes the absorption
(emission) of a photon by the system, while $\mathcal{L}_{0}$ collects
the terms that do not involve particle transfer. Here, we have assumed
that at most one photon is exchanged at a time.

To proceed, we introduce the counting field $\chi$ and the fourier-transform
density matrix:
\begin{equation}
\rho(\chi,t)=\sum_{N}\rho(N,t)\mathrm{e}^{-\mathrm{i}N\chi}.
\end{equation}
This leads to the generalized master equation \cite{Schaller2013,Kalaee2021}
\begin{equation}
\dot{\rho}(\chi,t)=\mathcal{L}_{\chi}\rho_{\chi}(t)\equiv\mathcal{L}\rho(\chi,t)+\sum_{k}\left(e^{i\nu_{k}\chi}-1\right)L_{k}\rho(\chi,t)L^{\dagger}_{k},
\end{equation}
where $\mathcal{L}$ denotes the Liouvillian defined in Eq. (3) of
the main text. In the limit $\chi\rightarrow0$, this reduces to the
original Liouvillian governing the standard time evolution. For the
three-level maser considered here, it is sufficient to monitor only
the quanta exchanged with bath $h$. Accordingly, we assign $\nu_{1}=-1$
and $\nu_{2}=1$, while $\nu_{3}=\nu_{4}=0$. For an initial state
$\rho_{0}$, assumed to be the steady state of $\mathcal{L}$, the
formal solution is $\rho(\chi,t)=e^{\mathcal{L}_{\chi}t}\rho_{0}$. 

The probability of $N$ photons being transferred into bath $c$ during
time $t$ is then given by
\begin{equation}
P_{N}(t)=\operatorname{Tr}[\rho(N,t)]=\frac{1}{2\pi}\int^{\pi}_{-\pi}\mathrm{d}\chi\operatorname{Tr}\left[\mathrm{e}^{\mathcal{L}_{\chi}t}\rho_{0}\right]\mathrm{e}^{\mathrm{i}N\chi}.
\end{equation}

From $P_{N}(t)$, the $k$-th moment follows as
\begin{equation}
\left\langle N^{k}(t)\right\rangle =\sum_{N}N^{k}P_{N}(t)=\left.\left(\mathrm{-i}\partial_{\chi}\right)^{k}\operatorname{Tr}\left[\rho(\chi,t)\right]\right|_{\chi=0}=\left.\left(\mathrm{-i}\partial_{\chi}\right)^{k}\operatorname{Tr}\left[\mathrm{e}^{\mathcal{L}_{\chi}t}\rho_{0}\right]\right|_{\chi=0}.
\end{equation}

While this expression provides direct access to the moments of the
distribution, it is analytically more convenient to work with cumulants.
To this end, we introduce the cumulant generating function
\begin{equation}
C(\chi,t)=\ln\operatorname{Tr}\left[\mathrm{e}^{\mathcal{L}_{\chi}t}\rho_{0}\right],
\end{equation}
which yields the cumulants of the distribution $P_{N}(t)$ via \cite{Kalaee2021},
\begin{equation}
\left\langle \left\langle N^{k}(t)\right\rangle \right\rangle =\left.\left(\mathrm{-i}\partial_{\chi}\right)^{k}C(\chi,t)\right|_{\chi=0}.
\end{equation}

The first two cumulants correspond to the mean and the variance, respectively:
\begin{equation}
\left\langle \left\langle N^{1}(t)\right\rangle \right\rangle =\langle N(t)\rangle,\quad\left\langle \left\langle N^{2}(t)\right\rangle \right\rangle =\left\langle N^{2}(t)\right\rangle -\langle N(t)\rangle^{2}=\operatorname{var}(N(t)).
\end{equation}
We are primarily interested in the steady-state behavior as $t\rightarrow\infty$.
If the system admits a unique stationary state, the generalized Liouvillian
$\mathcal{L}(\chi)$ has a single eigenvalue $\zeta(\chi)$. All other
eigenvalues acquire a strictly negative real part and thus decay away
at long times. Consequently, in the limit $t\rightarrow0$, the CGF
becomes approximately linear in time, i.e.,
\begin{equation}
C(\chi,t)=\ln\operatorname{Tr}\left[\mathrm{e}^{\mathcal{L}_{\chi}t}\rho_{0}\right]\simeq\ln\operatorname{Tr}\left[\mathrm{e}^{\zeta t}\rho_{0}\right]=\zeta(\chi)t.
\end{equation}

Since the cumulants grow linearly in time, it is often more informative
to consider their rates of growth rather than their absolute values.
Taking the time derivative of Eq. (30) yields
\begin{equation}
\left\langle \left\langle \dot{N}^{k}\right\rangle \right\rangle =\left.\left(-\mathrm{i}\partial_{\chi}\right)^{k}\zeta(\chi)\right|_{\chi=0}.
\end{equation}

In particular, the first derivative gives the mean current $J$ ,
while the second derivative yields the scaled varianc $D$ :
\begin{equation}
J=\left\langle \left\langle \dot{N}\right\rangle \right\rangle =-\left.i\partial_{\chi}\xi(\chi)\right|_{\chi=0},\quad D=\left\langle \left\langle \dot{N}^{2}\right\rangle \right\rangle =-\left.\partial^{2}_{\chi}\xi(\chi)\right|_{\chi=0}.
\end{equation}

As proven in Ref. \cite{Landi2024}, for the quantum jump unraveling,
the mean current $J$ and the noise $D$ admit explicit expressions:
\begin{equation}
J=\operatorname{Tr}\{\mathcal{J}\rho\}
\end{equation}
and
\begin{equation}
D=\sum_{k}\operatorname{Tr}\left\{ \nu^{2}_{k}L_{k}\rho L^{\dagger}_{k}\right\} -2\operatorname{Tr}\left\{ \mathcal{J}\mathcal{L}^{+}\mathcal{J}\rho\right\} ,
\end{equation}
respectively.

\section{The master equation of the quantum Scovil\textendash Schulz-DuBois
(SSDB) three-level maser}

For the quantum Scovil\textendash Schulz-DuBois (SSDB) three-level
maser, the Hamiltonian of the system is written as 
\begin{equation}
H(t)=H_{0}+V(t),
\end{equation}
with the bare contribution
\begin{equation}
H_{0}=\omega_{h}\sigma_{hh}+\omega_{c}\sigma_{cc}+\omega_{x}\sigma_{xx}
\end{equation}
and the external classical driving field
\begin{equation}
V(t)=\epsilon\left(\mathrm{e}^{\mathrm{i}\omega_{d}t}\sigma_{hc}+e^{-\mathrm{i}\omega_{d}t}\sigma_{ch}\right).
\end{equation}

To eliminate the explicit time dependence, we transform to an appropriate
rotating frame, which simplifies the equations of motion. Rotated
operators are defined via the unitary transformation
\begin{equation}
\tilde{X}=U^{\dagger}(t)AU(t),
\end{equation}
with $U(t)=\mathrm{e}^{\mathrm{-i}\bar{H}t}$ and $\bar{H}=\omega_{h}\sigma_{hh}+\left(\omega_{h}+\omega_{d}\right)\sigma_{cc}+\omega_{x}\sigma_{xx}$.
By assuming a spin-boson type system\textendash bath interaction and
making the Born\textendash Markov approximation, i.e., assuming that
the environment is stationary and its correlations decay much faster
than the system dynamics, the master equation in this rotating frame
takes the form \cite{Vu2022,Kalaee2021-1}
\begin{equation}
\frac{d}{dt}\rho=\mathcal{L}\rho=-i\left[H,\rho\right]+\sum_{k}\mathcal{D}\left(L_{k},\rho\right)
\end{equation}
with the effective Hamiltonian $H=-\Delta\sigma_{cc}+\epsilon\left(\sigma_{ch}+\sigma_{hc}\right)$
and the detuning parameter $\Delta=\omega_{d}+\omega_{h}-\omega_{c}$.
The jump operators $L_{1}=\sqrt{\gamma_{h}n_{h}}\sigma_{xh}$, $L_{2}=\sqrt{\gamma_{h}\left(n_{h}+1\right)}\sigma_{hx}$,
$L_{3}=\sqrt{\gamma_{c}n_{c}}\sigma_{xc}$, and $L_{4}=\sqrt{\gamma_{c}\left(n_{c}+1\right)}\sigma_{cx}$,
where the transition operators are defined as $\sigma_{ij}=\left|i\right\rangle \left\langle j\right|$.
Importantly, the structure of the dissipators remains unchanged under
this transformation.